# Contributions to the Theory of Thermostated Systems

Ronald F. Fox
Regents' Professor Emeritus
School of Physics
Georgia Institute of Technology
Atlanta, Georgia

## ABSTRACT

In this paper, a theory for systems in contact with a thermal reservoir is developed. We call such systems "thermostated systems." Interest in these systems has been vigorous for about the last 20 years and is illustrated by the Crooks theorem and the Jarzinski equality, two results for the description of systems in full phase space where all coordinates and momenta may be followed through time. These fundamental results follow from the behavior of Markovian systems in phase space for which path integral expressions can be derived from the Smoluchowski equation valid for Markovian systems. In the **Introduction** a brief account of this approach is given in which it is stressed that some of the motivation for the approach comes from computer simulations in which at an instant in time all momenta can be reversed in order to study the reversed trajectories. The impossibility of doing this reversal experimentally is discussed and a brief review of the origin of irreversibility in dynamical systems is given. Ultimately this leads to an alternative approach that involves "contraction of the description" and the derivation of a coordinate only picture that is intrinsically non-Markovian. In section **I** we explain how and why we begin our analysis with the Liouville-Langevin equation. The significance of the abundance of water in biological cells is presented and explains the appropriateness of the Liouville-Langevin equation. In section **II** the projection operator approach of Zwanzig and Mori, although in an essentially modified form, is developed. This shows why the results ultimately obtained are non-Markovian. In section **III** boson operator representations for the projection operator approach are derived. This makes the interpretation of the equations easier to grasp and facilitates the analysis. In section **IV** the intrinsic non-commutivity of the boson operators leads to time ordered exponentials in the general final result, equation (43), the central result of the paper. Section **V** contains an elementary check of equation (43) for the case of no potential energy term in the Liouville equation. In section **VI** what it means for the result to be non-Markovian is explained. Especially pertinent is the failure of the Smoluchowski equation in the non-Markovian case. Thus the path integral methods used in full phase space are not applicable in the contracted description case. Section **VIIa** contains an advanced check of equation (43) for the damped harmonic oscillator. The time evolution of the averaged coordinate is given. In section **VIIb** analysis of the memory kernel in equation (43) for the harmonic oscillator is presented in a quite long sequence of steps. The time evolution of the averaged square of the coordinate is derived. In section **VIII** the results for the harmonic oscillator are combined to form the conditional probability distribution for this non-Markovian, Gaussian example. In section **IX** non-equilibrium thermodynamics results are discussed. The demarcation between generalized non-equilibrium thermodynamic results governed by the Helmholtz free energy and non-thermodynamic results that feature the non-Markovian character of our general framework is elucidated. Underdamping is the key. Finally, in section **X** we address the incidence of underdamping in sub-cellular biology.

## Introduction

For the past twenty or so years, there has been a renewed interest in thermostated systems (e. g. [1] Evans et al., [2] Gallavotti and Cohen, [3] Crooks, [4] Jarzynski, [5] Gore et al., [6] Hoover and Hoover, [7] England). Some of this interest arose from the ability to do detailed molecular dynamics calculations in which it is possible to specify and reverse all momenta simultaneously and instantaneously. Markovian analysis of such systems gave rise to the Crooks theorem and to the Jarzynski equality, to name just two of the main results. Part of the interest stemmed from nano-scale atomic force microscopy experiments and from laser tweezers experiments in which it is possible to prepare macromolecules in initial states and observe their subsequent relaxation to thermal equilibrium.

Thermostated systems are the natural setting for nano-biology since activity at the sub-cellular level is effectively thermostated primarily by liquid water, the dominate constituent of cells, that has very good heat capacity and thermal conductivity properties, and in which all other cellular molecules are immersed. The water molecules communicate their thermostating effects to immersed molecules by Brownian motion, that is extremely robust at the nanometer scale. At this scale , dynamics is at very low Reynolds number, i.e. viscosity overwhelms inertia ([8] Fox, and references therein).

Our interest in thermostated systems began in the 1970s. We observed ([9] Davis and Fox) that the Liouville-Langevin evolution operator could be treated by operator algebra techniques that demonstrated that an initial, rapid stage of momentum relaxation to Maxwell's distribution, caused by Brownian motion, is followed by a much slower relaxation of the coordinate distribution to Boltzmann's distribution in a macroscopic system. At that time the explicit details of the second, slower relaxation, were not put into evidence, only the asymptotic final Boltzmann distribution was justified. Stimulated by the more recent interest of other researchers in thermostated systems, we have returned to this problem and have succeeded in advancing our earlier program.

Several approaches to thermostated systems exist in the literature. In many of these both coordinates and conjugate momenta are explicitly kept in evidence for all times. This is significantly the result of computer simulations that provide the opportunity to control the details in full phase space. However, in atomic force microscopy and laser tweezers experiments only coordinate control is possible because momenta are rapidly changing in liquid water on a time scale of around $10^{-13}$ seconds at room temperature $\left(\frac{h}{k_B T}\right)$. No possibility exists to simultaneously control all relevant momenta, say to reverse all of them at some instant, as is possible in computer simulations. When describing what is of interest in such a context, *coarse-graining* is sometimes invoked (e. g. [7] England, [10] Spinney and Ford). Coarse-graining arose long ago in order to help Boltzmann attempt to combat the assault on his ideas by Zermelo and others ([11] Fox, pp. 262-265, [12] Uhlenbeck).

The basic obstacle faced when explaining the origin of irreversibility in many particle, time reversible Newtonian systems is the Poincare recurrence theorem. In a spatially bounded (ergodic) system the Newtonian trajectory passes through every finite subvolume of phase space infinitely often. Boltzmann had argued that phase space was coarse-grained into many, but a finite number of, regions, the overwhelmingly largest of which was, for him, the equilibrium state. He argued that any trajectory starting in one of the small non-equilibrium regions would inevitably find its way into the very large equilibrium region where it would spend most of the time, in proportion to the size of the region. Zermelo quoted Poincare's recurrence theorem that said the trajectory would return to the initial non-equilibrium region infinitely often, thereby destroying any impression of irreversibility. This proved devastating to Boltzmann (suicide by hanging at age 62 in 1906, possibly because of an undiagnosed bipolar disorder) even though he argued that these returns would occur very briefly and represented the fluctuations in the macroscopic system.

It was Gibbs who began to understand how Boltzmann could have won the argument ([12] Uhlenbeck). Gibbs introduced ensembles of points in phase space. This is related to coarse-graining in that all the points in an initial non-equilibrium coarse-grained region could be taken as a Gibbs ensemble. Gibbs would start with a compact set of points in some small non-equilibrium region and let Newton's equations of motion dictate the subsequent trajectories. Now each and every ensemble point trajectory would eventually re-enter the initial region but they would do so at vastly different times as the ensemble of points became very filamentous in phase space with time, so that parts of it were in many different regions of phase space at any instant of time (ergodicity). On average over the ensemble, by far most point trajectories at any instant would be in the large equilibrium region of phase space. Those point trajectories outside the equilibrium region created the fluctuations observed around equilibrium. The Boltzmann-Gibbs picture was a real advance but not the whole story. What determines the size and shape of the different coarse-grained regions of phase space? As George Uhlenbeck used to say ([12] Uhlenbeck, p. 15), the coarse-graining, or the description of the Gibbs ensembles, depended on the "zeal of the observer." Thus it is subjective, and the theory of irreversibility is still not objective.

Uhlenbeck also saw a way out of this subjectivity. It is called "contraction of the description." ([13] Fox and Uhlenbeck I. and II., [14] Keizer, chapter 9). It has its roots in the work of Chapman and Enskog ([12] Uhlenbeck, Chapter VI). An example is the reduction of the many particle Newtonian phase space description to a hydrodynamic description (Navier-Stokes) in which there are just five densities, mass, energy, momentum (3-d) corresponding to the basic dynamically conserved quantities. This apparently removes the subjectivity although a strictly rigorous derivation of this claim for hydrodynamics is still not known. The efficacy of the Navier-Stokes hydrodynamic description in real practical situations such as movement of ships and submarines through water and the aviation of planes through air provides substantial justification for its validity.

In the present work we will achieve contraction by exactly integrating over all momenta, yielding a contracted description in spatial coordinates alone. The motivation for this approach is the realization that our visual observations, even using microscopes, see the movement of matter in space, but not the momenta of the individual molecules. As already mentioned the momenta change direction extremely frequently because of molecular collisions. In cellular biology most of the molecules are water and as such not readily visible because water is transparent to visible light. They are the carriers of heat to and from all molecules through induced Brownian motion of other molecules. Thus, our contraction will be from coordinates and momenta of all molecules (including water) to only the coordinates (excluding water). An example of this motivation is the observed self-assembly of a protein complex by ultramicroscopy in which the protein assembly is seen in space, paradoxically in the direction of decreasing entropy but where the greater, compensating increase of entropy by the associated water molecules goes unseen (to do a proper theory of this process it is necessary to explicitly include initially protein bound water molecules). To do this contraction for a thermostated system we first extend the Liouville equation to the Liouville –Langevin equation (sometimes called the Kramers equation) ([15] Chandrasekhar, chapter 2, section 4) that incorporates the thermostating and eliminates explicit dependence on the water molecules, and then integrate out the momenta to achieve contraction of the description to the coordinate variables of all other molecules. In the present analysis we do not start from pure Newtonian dynamics, as is done in the full phase space treatment of thermostats, but instead assume that the inclusion of the Langevin operator already has been justified by a first contraction from many particle dynamics using the projection operator techniques to be introduced below. For simplicity of presentation, the account here will be for a single particle in one dimension. Generalization to many particles in three dimensions is straight-forward although relatively cumbersome.

## I. Liouville-Langevin equation

The Liouville equations describes the evolution of a normalized, positive density, $D(r,p,t)$, in phase space that vanishes for infinite momenta and on the spatial boundaries. It satisfies the partial differential equation:

(1)
$$\partial_t D(r,p,t) = \left(-\frac{p}{m}\partial_r + U'(r)\partial_p\right) D(r,p,t)$$

where $-U'(r)$ is the force, and $D$ is non-negative and normalized:

(2)
$$D(r,p,t) \geq 0 \text{ and } \int dr \int dp\, D(r,p,t) = 1$$

in which the integration is over all of phase space. If $D$ vanishes at the boundaries a simple integration of the right-hand side of (1) shows that normalization is preserved for all times. This picture is equivalent to Newtonian dynamics represented in phase space. To incorporate a thermostat we add an evolution operator called the Langevin operator to the right-hand side. It represents Brownian motion and relaxes the momentum part of the distribution to a Maxwellian distribution on a time scale $m/\alpha$ where $m$ is the mass of the particle and $\alpha = 6\pi\eta R$ is the Stokes formula for the drag, $\alpha$, on the particle (taken to be a sphere, although the generalization to ellipsoids is known ([8] Fox, chapter 2)) of radius $R$ in a fluid of viscosity $\eta$, intended in the Boussinesq sense. The Langevin operator is:

(3)
$$\alpha \partial_p \left( \frac{p}{m} + \frac{1}{\beta} \partial_p \right)$$

where

$$\beta = \frac{1}{k_B T}$$

in which $T$ is the temperature and $k_B$ is Boltzmann's constant. The Liouville-Langevin equation is:

(4)
$$\partial_t D(r,p,t) = \left( -\frac{p}{m}\partial_r + U'(r)\partial_p + \alpha\partial_p\left(\frac{p}{m} + \frac{1}{\beta}\partial_p\right) \right) D(r,p,t)$$

The three operators, $A \equiv \frac{p}{m}\partial_r$, $U'(r)\partial_p$, and $B \equiv \alpha\partial_p\left(\frac{p}{m} + \frac{1}{\beta}\partial_p\right)$ do not commute so that the formal solution to (4) given by:

(5)
$$D(r,p,t) = \exp[t(-A + U'(r)\partial_p + B)]D(r,p,t)$$

is non-trivially complex. $\exp[-tA]$ is a $r$-translation by an amount that depends on $p$. $\exp[tU'(r)\partial_p]$ is a $p$-translation by an amount that depends on $r$ and $\exp[tB]$ is a type of momentum diffusion. Separately, each of these actions is easily rendered, but the exponential of their sum is not. In ([9] Davis and Fox) we used the three operators' commutator algebra to go as far as is possible with the algebra approach. Because of the presence of $U'(r)\partial_p$, the commutator algebra does not close as a finite algebra (except for the simple harmonic oscillator case), even though the $(A, B)$ commutator algebra does.

Suppose you have

(6)
$$exp[t(-A + B)]$$

where $[A, B] \neq 0$, as is the case here. In this case where $U'(r)\partial_p$ is absent, the commutator algebra for $A$ and $B$ creates two other operators: $C \equiv \frac{2}{\beta}\partial_p\partial_r$ and $D \equiv -\frac{2}{\beta m}\partial_r^2$, so that a finite closed algebra is obtained.

(7)
$$[A, B] = -\frac{\alpha}{m}(A + C) , [B, C] = -\frac{\alpha}{m}C , [A, C] = D$$
and
$$[A, D] = [B, D] = [C, D] = 0$$

This is what we used in ([9] Davis and Fox).

Suppose the initial state is given by

(8)
$$D(r, p, 0) = W_M(p)R(r, 0) \text{ with } W_M(p) \equiv \left(\frac{\beta}{2\pi m}\right)^{1/2} exp\left[-\frac{\beta}{2m}p^2\right]$$

where $W_M$ is the Maxwell distribution and $R(r, 0)$ is normalized, positive but otherwise arbitrary, then from the commutator algebra we get exactly the contraction of the description

(9)
$$\partial_t R(r, t) = D_E \left(1 - exp\left[-t\frac{\alpha}{m}\right]\right)\partial_r^2 R(r, t)$$
where
$$R(r, t) \equiv \int dp D(r, p, t)$$
and
$$D_E \equiv \frac{1}{\beta\alpha}$$

The last definition is Einstein's formula for the diffusion constant and is produced by the contraction. The time dependence in this non-Markovian generalization of the standard Markovian diffusion equation generates the Ornstein-Furth formula ([16] Uhlenbeck and Ornstein) for the mean square displacement for both long times and for short times compared to the Langevin time scale $m/\alpha$. Note that although the initial phase space distribution is factored into a $R$-part and a $W_M$-part, the $r$ and $p$ dependence becomes intertwined by the dynamics for all $t > 0$. This is the result we desire to generalize for the inclusion of the potential energy term $U(r)$.

## II. Projection Operators

We will use a modification of the Zwanzig-Mori ([17] Zwanzig, [18] Mori) projection operator technique ([11] Fox). Let $F$ and $G$ be arbitrary functions of $p$ and $r$. Focus on the $p$-dependence. Let the inner product in p-space be denoted by and defined by

(10)
$$\langle F|G\rangle \equiv \int dp\, W_M^{-1} F^* G$$

Note that unlike Zwanzig-Mori we choose the weight function to be the inverse of the Maxwell distribution. The projection operator, $P$, is defined by

(11)
$$PG \equiv \left(\int dp\, W_M^{-1} W_M G\right) W_M = |W_M\rangle\langle W_M|G\rangle$$

so that

$$PD \equiv \left(\int dp\, W_M^{-1} W_M D\right) W_M = |W_M\rangle\langle W_M|D\rangle = |W_M\rangle R(r,t)$$

This operator takes the phase space distribution, $D$, into the product of the contracted spatial distribution, $R$, times the Maxwell distribution, $W_M$. (The last equality in Eq.(11) follows from the second line of Eq.(9).)

Referring to Eq.(4) the Liouville operator is

(12)
$$L_0 = -A + U'\partial_p$$

and the Liouville-Langevin operator is

(13)
$$L = L_0 + B$$
where
$$BW_M = 0$$

Because $D(t = 0) = R(r, 0)W_M(p)$

(14)
$$(1 - P)D(t = 0) = 0$$
since
$$PD(t = 0) = D(t = 0)$$

Therefore

(15)
$$P^2 D(t) = P|W_M\rangle\langle W_M|D(t)\rangle = |W_M\rangle\langle W_M|D(t)\rangle$$
because
$$P|W_M\rangle = |W_M\rangle$$
and
$$\langle W_M|W_M\rangle = 1$$

given the choice of weight function for the inner product. In general $P$ has the properties

(16)
$$P^2 = P,$$
$$(1 - P)^2 = 1 + P^2 - 2P = 1 - P,$$
$$P(1 - P) = 0$$

that characterize orthogonal projection operators $P$ and $1 - P$.

Starting from Eq.(4) in the form $\partial_t D = LD$ we get the projection equations

(17)
$$\partial_t PD = PLPD + PL(1 - P)D$$
and
$$\partial_t (1 - P)D = (1 - P)L(1 - P)D + (1 - P)LPD$$

The formal solution to the second equation is

(18)
$$(1 - P)D(t) = exp[t(1 - P)L(1 - P)](1 - P)D(0)$$
$$+ \int_0^t ds\, exp[(t - s)(1 - P)L(1 - P)](1 - P)LPD(s)$$

wherein Eq.(16) has been used for $1-P$ in both lines. The initial value term that is the first line vanishes because of Eq.(14). Substitution into the first line of Eq.(17) yields

(19)
$$\partial_t PD = PLPD + PL \int_0^t ds\, exp[(t-s)(1-P)L(1-P)](1-P)LPD(s)$$

This contracted equation can be simplified significantly. From the second lines of Eq.(11) and Eq.(13) it follows that

(20)
$$LPD(t) = (L_0 + B)R(r,t)W_M(p) = L_0 R(r,t)W_M(p)$$

Therefore, it follows that

(21)
$$PLPD = \left(\int dp\, W_M^{-1} W_M L_0 R W_M\right) W_M = 0$$

because
$$L_0 R(r,t)W_M = \left(-\frac{p}{m}\partial_r - \frac{\beta}{m}U'p\right)R(r,t)W_M$$

and
$$\int dp\, W_M^{-1} W_M p W_M = 0$$

where the last line follows from integrating the product of an odd function of $p$ and an even function of $p$ over a symmetric $p$-domain. Clearly, for arbitrary $G$ this implies

(22)
$$PLPG = PLW_M \langle W_M | G \rangle = PL_0 W_M \langle W_M | G \rangle = 0$$

Thus, any occurrence of the structure $PLP$ makes a term vanish. The projection equation, Eq.(19) becomes

(23)
$$\partial_t PD = PL \int_0^t ds\, exp[(t-s)(1-P)L(1-P)]LPD(s)$$

This is manifestly non-Markovian as a result of the contraction of the description and as is manifested by the time convolution integral. It is the basic equation from which all othe results follow including the central result in Eq.(43). It was used originally by Mori to attempt to justify the Langevin equation. Instead of obtaining the Markovian Langevin equation he obtained the "generalized Langevin equation" that is non-Markovian ([11] Fox, sections I.3 and I.4).

## III. Boson Operator Representation

The $r$-dependent analog to the Maxwell distirbution is the Boltzmann distribution, $W_B = (1/Q)exp[-\beta U(r)]$ in which $Q = \int dr\, exp[-\beta U(r)]$. The $r$-dependent inner product is defined by

(24)
$$\langle f|g\rangle \equiv \int dr\, W_B^{-1} f^* g$$

in which we have again used the inverse distribution for the weight function. Define boson operators $a$ and $a^\dagger$ for the inner product in Eq.(10) by

(25)
$$a = -\sqrt{\frac{m}{\beta}}\, \partial_p - \sqrt{\frac{\beta}{m}}\, p$$

and

$$a^\dagger = \sqrt{\frac{m}{\beta}}\, \partial_p$$

These operators are dimensionless and satisfy the commutation relation

(26)
$$[a, a^\dagger] = 1$$

Define boson-like operators $b$ and $b^\dagger$ for the inner product in Eq.(24) by

(27)
$$b = -\sqrt{\frac{\beta}{m}}\left(\frac{1}{\beta}\partial_r + U'\right)$$

and

$$b^\dagger = \sqrt{\frac{1}{m\beta}}\, \partial_r$$

These operators have the dimensions of inverse seconds and satisfy the commutation relation

(28)
$$[b, b^\dagger] = \frac{U''}{m}$$

Only for the special case of the harmonic oscillator potential, $U = \frac{1}{2}m\omega^2 r^2$, are these operators proportional to genuine boson operators, and the right-hand side of Eq.(28) becomes $\omega^2$. We could divide $b$ and $b^\dagger$ by $\omega$ to make them dimensionless, but only for the harmonic oscillator potential does the commutator algebra close after one iteration, like for genuine boson operators. Keeping the dimensions as is will serve as a reminder of this limitation.

The $a, a^\dagger$ ground state is determined solely by the commutation relation and the inner product. We may write

(29)
$$|W_M(p)\rangle \equiv |0\rangle$$
and
$$\langle 0|0\rangle = 1$$
since
$$a|W_M(p)\rangle = 0$$

Therefore we also have the canonical identities

(30)
$$a^\dagger|0\rangle = |1\rangle$$
$$(a^\dagger)^n|0\rangle = \sqrt{n!}\,|n\rangle$$
$$\langle 0|(a)^n = \sqrt{n!}\,\langle n|$$

What is the expression for the projection operator, $P$, in terms of the boson operators? The answer is

(31)
$$P = |0\rangle\langle 0| \otimes 1$$

a direct product of the projection onto $W_M(p)$ and the identity in $r$-space. Note that we can write

(32)
$$B = -\frac{\alpha}{m} a^\dagger a$$
and

$$L_0 = b^\dagger a - b a^\dagger$$

For Eq.(23) we need

(33)
$$(1-P)L(1-P)$$
$$= (1 - |0\rangle\langle 0| \otimes 1)\left(b^\dagger a - ba^\dagger - \frac{\alpha}{m}a^\dagger a\right)(1 - |0\rangle\langle 0| \otimes 1)$$
$$= b^\dagger(a - |0\rangle\langle 1|) - b(a^\dagger - |1\rangle\langle 0|) - \frac{\alpha}{m}a^\dagger a$$

because
$$a|0\rangle = 0$$
and
$$\langle 0|a^\dagger = 0$$

Moreover, it follows similarly for arbitrary $G$ that

(34)
$$PLG = |0\rangle\langle 0|\left(b^\dagger a - ba^\dagger - \frac{\alpha}{m}a^\dagger a\right)G\rangle = |0\rangle\langle 1|b^\dagger G\rangle$$

and
$$LPD(s) = \left(b^\dagger a - ba^\dagger - \frac{\alpha}{m}a^\dagger a\right)|0\rangle\langle 0|D(s)\rangle = -b|1\rangle\langle 0|D(s)\rangle$$

Using $\langle 0|D(s)\rangle = \int dp\, W_M^{-1} W_M D(r,p,s) = R(r,s)$ and Eq.(34) we conclude that

(35)
$$\partial_t R(r,t)$$
$$= -b^\dagger\langle 1|\int_0^t ds\, \exp\left[(t-s)\left(b^\dagger(a - |0\rangle\langle 1|) - b(a^\dagger - |1\rangle\langle 0|)\right.\right.$$
$$\left.\left. - \frac{\alpha}{m}a^\dagger a\right)\right]|1\rangle b R(r,s)$$

The expectation value over the first excited state in the momentum distribution amounts to integrating out all momentum dependence and leaving a reduced (contracted) distribution in the spatial variable only. The expression above retains an operator character in the $r$-variable. It may be noted immediately from the definition in Eq.(27) that the Boltzmann distribution is a stationary solution of this equation because $bW_B(r) = 0$. It may be demonstrated that the Boltzmann distrbution is the unique equilibrium distribution to which any initial distribution is driven by Eq.(35). We will leave this matter until later, after we have developed more manageable representations of the equation.

# IV. Non-commutivity and Time Ordered Exponentials

The convolution kernel in Eq.(35) is the exponential of the sum of three mutually non-commuting terms. Because of the commutation identities

(36)
$$[a^\dagger a, (a - |0\rangle\langle 1|)] = -(a - |0\rangle\langle 1|)$$
and
$$[a^\dagger a, (a^\dagger - |1\rangle\langle 0|)] = (a^\dagger - |1\rangle\langle 0|)$$

the kernel can be written

(37)
$$exp\left[(t-s)\left(b^\dagger(a - |0\rangle\langle 1|) - b(a^\dagger - |1\rangle\langle 0|) - \frac{\alpha}{m}a^\dagger a\right)\right]$$
$$= exp\left[-(t-s)\frac{\alpha}{m}a^\dagger a\right] \times$$
$$\overleftarrow{T}exp\left[\int_s^t ds'\left(b^\dagger(a - |0\rangle\langle 1|)exp\left[-\frac{\alpha}{m}(s'-s)\right]\right.\right.$$
$$\left.\left. - b(a^\dagger - |1\rangle\langle 0|)exp\left[\frac{\alpha}{m}(s'-s)\right]\right)\right]$$

In this expression we have the time ordered exponential defined by

(38)
$$\overleftarrow{T}exp\left[\int_s^t ds'[O(s')]\right] \equiv 1 + \sum_{n=1}^\infty \int_s^t ds_1 \int_s^{s_1} ds_2 \ldots \int_s^{s_{n-1}} ds_n\, O(s_1)O(s_2)\ldots O(s_n)$$

in which $t \geq s_1 \geq s_2 \geq \cdots \geq s_n \geq s$ and the operstor, $O(s)$, does not commute with itself for two different times. This is the left-ward ordered exponential with later times to the left of earlier times. It satisfies the first order differential identity

(39)
$$\frac{d}{dt}\overleftarrow{T}exp\left[\int_s^t ds'[O(s')]\right] = O(t)\overleftarrow{T}exp\left[\int_s^t ds'[O(s')]\right]$$

Especially note that the $O$ out front on the right hand side is at time $t$ and is on the left. The presence of the factors $exp\left[\mp\frac{\alpha}{m}s'\right]$ in Eq.(37) come about in a manner similar to the simplified rendering below;

(40)
$$exp\left[(t-s)\left(\mu a - \frac{\alpha}{m}a^\dagger a\right)\right] = exp\left[-(t-s)\frac{\alpha}{m}a^\dagger a\right] \times$$

$$\overleftarrow{T}exp\left[\int_s^t ds' \mu \, exp\left[(s'-s)\frac{\alpha}{m}a^\dagger a\right] a \, exp\left[-(s'-s)\frac{\alpha}{m}a^\dagger a\right]\right]$$

Using Eq.(39) the $t$-derivative of both sides is the same, and both sides agree for $t = s$, sufficient conditions for a unique solution to a first order equation. The exponent factor on the right-hand side may be expressed by a commutator operator as

(41)
$$exp\left[(s'-s)\frac{\alpha}{m}a^\dagger a\right] a \, exp\left[-(s'-s)\frac{\alpha}{m}a^\dagger a\right] = exp\left[(s'-s)\frac{\alpha}{m}[a^\dagger a, \bullet]\right] a$$
$$= a \, exp\left[-(s'-s)\frac{\alpha}{m}\right]$$

in which the the commutator operator, $[a^\dagger a, \bullet]$, is defined by $[a^\dagger a, \bullet]O = [a^\dagger a, O]$ and the powers of this commutator implicit in its exponential above are just the iterated commutators, and we have used Eq.(36).

We may now use Eq.(37) in Eq.(35) and let the leading exponential factor act leftwards on the state $\langle 1|$, i.e.

(42)
$$\langle 1| exp\left[-(t-s)\frac{\alpha}{m}a^\dagger a\right] = \langle 1| exp\left[-(t-s)\frac{\alpha}{m}\right]$$

yielding the contracted equation (our central result)

(43)
$$\partial_t R(r,t) = -b^\dagger \langle 1| \int_0^t ds \, exp\left[-(t-s)\frac{\alpha}{m}\right] \times$$
$$\overleftarrow{T}exp\left[\int_s^t ds' \left(b^\dagger(a - |0\rangle\langle 1|)exp\left[-\frac{\alpha}{m}(s'-s)\right]\right.\right.$$
$$\left.\left. - b(a^\dagger - |1\rangle\langle 0|)exp\left[\frac{\alpha}{m}(s'-s)\right]\right)\right]$$
$$\times |1\rangle b R(r,s)$$

It might be thought that for the harmonic oscillator potential we have an exactly solvable dynamics. However the coupling of the effective Langevin oscillator operators with the real oscillator operators is multiplicative rather than additive. This renders the two-coupled-oscillator problem insolvable as has been noted by Louisell, ([19] Louisell). Nevertheless, for the contracted description given here the oscillator case will be seen to be tractable and will later serve as a check of the validity of the central result given in Eq.(43).

## V. Elementary Checks of Eq.(43)

The minimal check is for the case $U=0$. This makes $b = -b^\dagger$ (see Eq.(27). Eq.(43) in turn becomes
(44)
$$\partial_t R(r,t) = b^\dagger \int_0^t ds\ exp\left[-(t-s)\frac{\alpha}{m}\right] \times$$
$$\langle 1|\overleftarrow{T} exp\left[\int_s^t ds' b^\dagger \left((a - |0\rangle\langle 1|)exp\left[-\frac{\alpha}{m}(s'-s)\right]\right.\right.$$
$$\left.\left. + (a^\dagger - |1\rangle\langle 0|)exp\left[\frac{\alpha}{m}(s'-s)\right]\right)\right]|1\rangle$$
$$\times b^\dagger R(r,s)$$

If we expand the time ordered exponential according to the definition in Eq.(38) the $\langle 1|...|1\rangle$ expectation values are non-zero for even order terms only, since they must include equal amounts of $a$ and $a^\dagger$. That means the $\langle 1|...|1\rangle$ expectation value of the time ordered exponential is an even order only power series in $b^\dagger$ with time dependent coefficients. The first few terms are (using Eq.(30) and Eq.(33))
(45)
$$\langle 1|\overleftarrow{T} exp\left[\int_s^t ds' b^\dagger \left((a - |0\rangle\langle 1|)exp\left[-\frac{\alpha}{m}(s'-s)\right]\right.\right.$$
$$\left.\left. + (a^\dagger - |1\rangle\langle 0|)exp\left[\frac{\alpha}{m}(s'-s)\right]\right)\right]|1\rangle$$
$$= 1 + 2b^\dagger b^\dagger \int_s^t ds_1 \int_s^{s_1} ds_2\ exp\left[-\frac{\alpha}{m}(s_1 - s_2)\right] +$$
$$\left(2^2 (b^\dagger b^\dagger)^2 + 3!(b^\dagger b^\dagger)^2\right)\int_s^t ds_1 ... \int_s^{s_3} ds_4\ exp\left[-\frac{\alpha}{m}(s_1 - s_2 + s_3 - s_4)\right] ...$$

The complexity of the terms in the series grows rapidly with the order of the term. This results from the different orderings of equal amounts of $(a - |0\rangle\langle 1|)$ and $(a^\dagger - |1\rangle\langle 0|)$. For example, in the above expression we have explicitly the second and fourth order terms for $U = 0$. The corresponding coefficient for the sixth order term (containing $(b^\dagger b^\dagger)^3$ and an eight-fold time integral) works out to be

(46)
$$2^3 + 4! + 2 \times 3^2 + 3 \times 2^2 + 2^2 \times 3$$

Consider calculating the moments of $r$. Multiply Eq.(44) from the left by $r$ and integrate by parts over $r$. Because of the two factors of $b^\dagger$ outside the time ordered exponential, the result vanishes.

(47)
$$\partial_t \langle r \rangle = \int dr\, r \partial_t R = 0$$

Consequently $\langle r \rangle = r_0$ for all time. Choose the origin of coordinates to be $r_0 = 0$. Now do the same for $r^2$. Only the 1 in the time ordered exponential expansion yields a non-zero result

(48)
$$\partial_t \langle r^2 \rangle = \int dr\, r^2 \partial_t R(r, t) = \int dr\, r^2 b^\dagger b^\dagger \int_0^t ds\, \exp\left[-(t-s)\frac{\alpha}{m}\right] R(r, s)$$
$$= 2 \frac{1}{m\beta} \frac{m}{\alpha} \left(1 - \exp\left[-t\frac{\alpha}{m}\right]\right)$$
$$= 2 D_E \left(1 - \exp\left[-t\frac{\alpha}{m}\right]\right)$$

This is precisely the same result obtained in the same way from Eq.(9). The integral of this equation is the Ornstein-Furth equation ([16] Uhlenbeck and Ornstein)

(49)
$$\langle r^2 \rangle_t = 2 D_E \left(t + \frac{m}{\alpha}\left(\exp\left[-t\frac{\alpha}{m}\right] - 1\right)\right)$$

Using Eq.(9) we can also get $\langle r^4 \rangle_t$ and check it against the Gaussian moment property that says

(50)
$$\langle r^4 \rangle_t = 3 \langle r^2 \rangle_t \times \langle r^2 \rangle_t$$

Eq.(9) implies

(51)
$$\partial_t \langle r^4 \rangle_t = \int dr\, r^4 \partial_t R(r, t) = 12 D_E \left(1 - \exp\left[-t\frac{\alpha}{m}\right]\right) \langle r^2 \rangle_t$$

Note that the moments on the left and right hand sides are evaluated at time $t$. Putting Eq.(49) into Eq.(51) verifies Eq.(50). This result also follows from Eq.(44) although in a rather different way. Now we must keep the second term in the expansion of Eq.(45).

(52)
$$\partial_t \langle r^4 \rangle_t = \int dr\, r^4 \partial_t R(r,t)$$
$$= \int dr\, r^4 b^\dagger b^\dagger \int_0^t ds\, \exp\left[-(t-s)\frac{\alpha}{m}\right]\left(1 \right.$$
$$\left. + 2 b^\dagger b^\dagger \int_s^t ds_1 \int_s^{s_1} ds_2 \exp\left[-\frac{\alpha}{m}(s_1-s_2)\right]\right) R(r,s)$$
$$= 12 \frac{1}{m\beta} \int_0^t ds\, \exp\left[-(t-s)\frac{\alpha}{m}\right]\left(\langle r^2 \rangle_s \right.$$
$$\left. + 2 \frac{1}{m\beta} \int_s^t ds_1 \int_s^{s_1} ds_2 \exp\left[-\frac{\alpha}{m}(s_1-s_2)\right]\right)$$
$$= 12 \frac{1}{\alpha\beta} \int_0^t ds\, \frac{\alpha}{m} \exp\left[-(t-s)\frac{\alpha}{m}\right]\left(\langle r^2 \rangle_s \right.$$
$$\left. + 2 \frac{1}{m\beta}\frac{m}{\alpha}\left(t-s+\frac{m}{\alpha}\left(\exp\left[-(t-s)\frac{\alpha}{m}\right]-1\right)\right)\right)$$
$$= 24 D_E^2 \int_0^t ds\, \frac{\alpha}{m} \exp\left[-(t-s)\frac{\alpha}{m}\right]\left(\left(s+\frac{m}{\alpha}\left(\exp\left[-s\frac{\alpha}{m}\right]-1\right)\right)\right.$$
$$\left. + 2\left(t-s+\frac{m}{\alpha}\left(\exp\left[-(t-s)\frac{\alpha}{m}\right]-1\right)\right)\right)$$
$$= \frac{d}{dt} 3(\langle r^2 \rangle_t)^2$$

wherein the last equality requires finishing the integrals. Note that $\langle r^2 \rangle_s$ appeared part way through, at time $s$ and not at time $t$.

One could proceed to calculate higher order moments by expanding the time ordered exponential to higher order. However, since diffusion is a Gaussian process the first two moments determine all higher order moments. From Eq.(49), using $D_E(t)$ to denote $D_E\left(t + \frac{m}{\alpha}\left(exp\left[-t\frac{\alpha}{m}\right] - 1\right)\right)$, the distribution function for $r$ is given by

(53)
$$W(r,t) = \frac{1}{\sqrt{4\pi D_E(t)}} exp\left[-\frac{r^2}{4D_E(t)}\right]$$

However, this is a non-Markovian generalization of diffusion because of the non-linear time dependence in $D_E(t)$.

## VI. What it Means to be non-Markovian

The distribution function Eq.(53) is actually the conditional probability distribution, given that the diffusing particle is initially at $r_0$ at time $t_0$ and is distributed at $r_1$ at time $t_1 > t_0$. We write Eq.(53) given these conditions as the two time conditional probability distribution ([20] Fox, [21] Wang and Uhlenbeck)

(54)
$$P_2(r_1, t_1; r_0, t_0) = \frac{1}{\sqrt{4\pi D_E(t_1 - t_0)}} exp\left[-\frac{(r_1 - r_0)^2}{4D_E(t_1 - t_0)}\right]$$

where

$$D_E(t_1 - t_0) = D_E\left((t_1 - t_0) + \frac{m}{\alpha}\left(exp\left[-(t_1 - t_0)\frac{\alpha}{m}\right] - 1\right)\right)$$

Note that

(55)
$$\lim_{t_1 \to t_0} P_2(r_1, t_1; r_0, t_0) = \delta(r_1 - r_0)$$

For Markovian diffusion, we have instead $D_E(t_1 - t_0) = \frac{1}{\beta\alpha} \times (t_1 - t_0)$. This is valid for times $(t_1 - t_0) \gg \frac{m}{\alpha}$. For the non-Markovian diffusion, $D_E(t_1 - t_0)$ is a nonlinear function of time whereas for the Markovian case it is a constant times $(t_1 - t_0)$ i.e. linear in time.

In the Markovian case, the Smoluchowski ([21] Wang and Uhlenbeck, note II of the appendix), or Chapman-Kolmogorov ([22] Arnold, chapter 2), equation is satisfied

(56)

$$P_2(r_2, t_2; r_0, t_0) = \int_{-\infty}^{\infty} dr\, P_2(r_2, t_2; r, s) P_2(r, s; r_0, t_0)$$

for $t_2 \geq s \geq t_0$. This equation in turn can be generalized to include many intermediate points (multi-time correlation distributions) ultimately leading to a path integral representation of the Markovian time evolution. Indeed, this is what has been done at the heart of the arguments made by Jarzinski and by Crooks ([4] Jarzynski and [3] Crooks). For the non-Markovian case given by Eq.(54), Eq.(56) is no longer valid. Indeed, the Doob theorem ([21] Wang and Uhlenbeck, [23] Doob) implies that the validity of Eq.(56) requires that $D_E(t_1 - t_0) = \frac{1}{\beta\alpha} \times (t_1 - t_0)$.

Even for the simple case of diffusion, events occurring on the time scale $\frac{m}{\alpha}$ or shorter times necessitate a non-Markovian description and do not permit the Markovian analysis and path integral techniques used in phase space by Jarzynski, Crooks and many others. In the general ($U \neq 0$) non-Markovian case the two time correlation distribution, $P_2(r_1, t_1; r_0, t_0)$, no longer determines all higher order multi-time correlations as it can do in the Gaussian Markovian case.

### VII.a Advanced Checks of Eq.(43)

The harmonic oscillator potential, $U = \frac{1}{2} m\omega^2 r^2$, provides a more penetrating look into Eq.(43). Multiplying from the left by $r$ and integrating by parts implies that the only contribution from the time ordered exponential is from the first term, 1, and the leading term of the right-hand side is $-rb^\dagger \to \frac{1}{\sqrt{m\beta}}$ yielding

(57)
$$d_t \langle r \rangle_t = \frac{1}{\sqrt{m\beta}} \int_0^t ds\, exp\left[-(t-s)\frac{\alpha}{m}\right] \langle b \rangle_s$$

$$= \frac{1}{\sqrt{m\beta}} \int_0^t ds\, exp\left[-(t-s)\frac{\alpha}{m}\right] \sqrt{\frac{\beta}{m}} (-m\omega^2 \langle r \rangle_s)$$

$$= -\omega^2 \int_0^t ds\, exp\left[-(t-s)\frac{\alpha}{m}\right] \langle r \rangle_s$$

Therefore, by differentiation, we get the standard equation for a damped harmonic oscillator

(58)
$$d_t^2 \langle r \rangle_t = -\frac{\alpha}{m} d_t \langle r \rangle_t - \omega^2 \langle r \rangle_t$$

Much more difficult to obtain is the equation for $\langle r^2 \rangle_t$. This is because $-r^2 b^\dagger \to \frac{2}{\sqrt{m\beta}} r$ and $r$ times the time ordered exponential creates non-vanishing terms from every order of the expansion of type Eq.(38). What we saw in Eq.(45) and Eq.(46) for $U = 0$, becomes for the harmonic oscillator the even order terms beginning with

(59)
$$2n = 2: \quad -2b^\dagger b \, exp\left[-(s_1 - s_2)\frac{\alpha}{m}\right]$$

$$2n = 4: \quad \left(2^2 (b^\dagger b)^2 exp\left[-(s_1 - s_2 + s_3 - s_4)\frac{\alpha}{m}\right]\right.$$
$$\left. + 3! \, b^\dagger b^\dagger bb \, exp\left[-(s_1 - s_3 + s_2 - s_4)\frac{\alpha}{m}\right]\right)$$

$$2n = 6: \quad \left(-2^3 b^\dagger b b^\dagger b b^\dagger b \, exp\left[-(s_1 - s_2 + s_3 - s_4 + s_5 - s_6)\frac{\alpha}{m}\right]\right.$$
$$- 3 \times 2^2 b^\dagger b^\dagger b b b^\dagger b \, exp\left[-(s_1 - s_3 + s_2 - s_4 + s_5 - s_6)\frac{\alpha}{m}\right]$$
$$- 3 \times 2^2 b^\dagger b b^\dagger b^\dagger b b \, exp\left[-(s_1 - s_2 + s_3 - s_5 + s_4 - s_6)\frac{\alpha}{m}\right]$$
$$- 2 \times 3^2 b^\dagger b^\dagger b b^\dagger b b \, exp\left[-(s_1 - s_3 + s_2 - s_5 + s_4 - s_6)\frac{\alpha}{m}\right]$$
$$\left. - 4! \, b^\dagger b^\dagger b^\dagger b b b \, exp\left[-(s_1 - s_4 + s_2 - s_5 + s_3 - s_6)\frac{\alpha}{m}\right]\right)$$

The complexity increases rapidly with the order index, $2n$. The order of the time variables is sequential for the first term only for each order. The variations are in evidence for $2n = 4$ and for $2n = 6$. What becomes clear upon further analysis is that for even order $2n$, there is always a term $(-2b^\dagger b)^n$ Moreover this is the only term that does not contain consecutive $b^\dagger$'s somewhere in the product and consecutive $b^\dagger$'s always come before consecutive $b$'s. The significance of these facts is that a factor of $r$ from the left of the time ordered exponential will create non-zero results for each $(-2b^\dagger b)^n$ factor and for no others. That $b$ factors cannot occur to the extreme left follows from the fact that they occur as $b(a^\dagger - |1 \rangle \langle 0|)$ and $\langle 1|b(a^\dagger - |1\rangle\langle 0|) = 0$. When integration by parts is done the effect on a term such as $r(b^\dagger b)^n$ is $r(b^\dagger b)^n \to \omega^{2n} r$. Putting all these pieces together in Eq.(43) yields

(60)
$$d_t \langle r^2 \rangle_t = \frac{2}{\sqrt{m\beta}} \int_0^t ds \; exp\left[-(t-s)\frac{\alpha}{m}\right] \times$$

$$\left(1 + \sum_{n=1}^\infty (-2\omega^2)^n \int_s^t ds_1 \int_s^{s_1} ds_2 \ldots \int_s^{s_{2n-1}} ds_{2n} \; exp\left[-(s_1 - s_2 + \cdots - s_{2n})\frac{\alpha}{m}\right]\right)$$

$$\times \left(\frac{1}{\sqrt{m\beta}} - \sqrt{\frac{\beta}{m}} m\omega^2 \langle r^2 \rangle_s\right)$$

If we define the middle bracket containing the multiple time integrals to be $I(t-s)$ then Eq.(60) can be written more compactly as

(61)
$$d_t \langle r^2 \rangle_t = 2 \int_0^t ds \; exp\left[-(t-s)\frac{\alpha}{m}\right] I(t-s) \left(\frac{1}{m\beta} - \omega^2 \langle r^2 \rangle_s\right)$$

The problem is reduced to evaluating $I(t-s)$.

### VII.b Evaluation of $I(t-s)$ and $\langle r^2 \rangle_t$

(If the reader prefers not to go through the details of this analysis, the final result is Eq.(94).)

The time derivative of $I(t-s)$ is straightforward

(62)
$$d_t I(t-s) = 0 +$$

$$d_t \sum_{n=1}^\infty (-2\omega^2)^n \int_s^t ds_1 \int_s^{s_1} ds_2 \ldots \int_s^{s_{2n-1}} ds_{2n} \; exp\left[-(s_1 - s_2 + \cdots - s_{2n})\frac{\alpha}{m}\right]$$

$$= -2\omega^2 \int_s^t ds_2 \; exp\left[-(t-s_2)\frac{\alpha}{m}\right] +$$

$$2^2 \omega^4 \int_s^t ds_2 \int_s^{s_2} ds_3 \int_s^{s_3} ds_4 \; exp\left[-(t - s_2 + s_3 - s_4)\frac{\alpha}{m}\right] +$$

$$\sum_{n=3}^{\infty}(-2\omega^2)^n \int_S^t ds_2 \int_S^{s_2} ds_3 \ldots \int_S^{s_{2n-1}} ds_{2n} \, exp\left[-(t-s_2+\cdots-s_{2n})\frac{\alpha}{m}\right]$$

Note the new location for $t$ inside the exponential as well as in the upper limit of the first integration. This leads to the equation and initial conditions

(63)
$$d_t^2 I(t-s) = -\frac{\alpha}{m} d_t I(t-s) - 2\omega^2 I(t-s)$$

with
$$I(t=s) = 1, \; d_t I(t+s) = 0 \; and \; d_t^2 I(t-s) = -2\omega^2$$

It is instructive to treat this second order equation in one variable as a first order equation in two. Introduce $J \equiv \frac{1}{\sqrt{2\omega}} d_t I$. $I$ and $J$ are dimensionless and $J$ satisfies $d_t J = -\frac{\alpha}{m} J - \sqrt{2}\omega I$ with initial condition $J(t=s) = 0$. Now write the two variable equations in matrix form

(64)
$$d_t \begin{pmatrix} I \\ J \end{pmatrix} = \begin{pmatrix} 0 & \sqrt{2}\omega \\ -\sqrt{2}\omega & -\frac{\alpha}{m} \end{pmatrix} \begin{pmatrix} I \\ J \end{pmatrix}$$

Let the matrix $M$ be defined by

(65)
$$M \equiv \begin{pmatrix} 0 & \sqrt{2}\omega \\ -\sqrt{2}\omega & -\frac{\alpha}{m} \end{pmatrix}$$

This matrix can be expressed in terms of the Pauli spin matrices

(66)
$$\sigma_0 = \begin{pmatrix} 1 & 0 \\ 0 & 1 \end{pmatrix} \; \sigma_z = \begin{pmatrix} 1 & 0 \\ 0 & -1 \end{pmatrix} \; \sigma_y = \begin{pmatrix} 0 & -i \\ i & 0 \end{pmatrix}$$

yielding

(67)
$$M = -\frac{\alpha}{2m}(\sigma_0 - \sigma_z) + i\sqrt{2}\omega\sigma_y$$

Using the initial conditions at $t = s$, the solution to Eq.(64) is
(68)

$$\begin{pmatrix} I(t-s) \\ J(t-s) \end{pmatrix} = \exp[(t-s)M]\begin{pmatrix} 1 \\ 0 \end{pmatrix}$$
$$= \exp\left[-(t-s)\frac{\alpha}{2m}\right]\exp\left[(t-s)\left(\frac{\alpha}{2m}\sigma_z + i\sqrt{2}\omega\sigma_y\right)\right]\begin{pmatrix} 1 \\ 0 \end{pmatrix}$$

because $\sigma_0$ commutes with the other two Pauli matrices and can be factored out in front. Any good introductory quantum mechanics book will have a justification of the identity

(69)
$$\exp\left[(t-s)\left(\frac{\alpha}{2m}\sigma_z + i\sqrt{2}\omega\sigma_y\right)\right]$$
$$= \text{Cosh}\left[(t-s)\sqrt{\frac{\alpha^2}{4m^2} - 2\omega^2}\right]\sigma_0$$
$$+ \frac{\text{Sinh}\left[(t-s)\sqrt{\frac{\alpha^2}{4m^2} - 2\omega^2}\right]}{\sqrt{\frac{\alpha^2}{4m^2} - 2\omega^2}}\left(\frac{\alpha}{2m}\sigma_z + i\sqrt{2}\omega\sigma_y\right)$$

Plugging this into Eq.(68) and reading off the upper component of the 2-vector gives

(70)
$$I(t-s) = \exp\left[-(t-s)\frac{\alpha}{2m}\right]\left(\text{Cosh}\left[(t-s)\sqrt{\frac{\alpha^2}{4m^2} - 2\omega^2}\right]\right.$$
$$\left. + \frac{\text{Sinh}\left[(t-s)\sqrt{\frac{\alpha^2}{4m^2} - 2\omega^2}\right]}{\sqrt{\frac{\alpha^2}{4m^2} - 2\omega^2}}\frac{\alpha}{2m}\right)$$

This may be placed inside Eq.(61). The product in the kernel can be rendered as a sum of exponentials

(71)
$$\exp\left[-(t-s)\frac{\alpha}{m}\right]I(t-s) =$$

$$\frac{1}{2}\left(\left(1+\frac{\alpha}{2m}\frac{1}{\sqrt{\frac{\alpha^2}{4m^2}-2\omega^2}}\right)\exp\left[-(t-s)\frac{3\alpha}{2m}+(t-s)\sqrt{\frac{\alpha^2}{4m^2}-2\omega^2}\right]\right.$$

$$\left.+\left(1-\frac{\alpha}{2m}\frac{1}{\sqrt{\frac{\alpha^2}{4m^2}-2\omega^2}}\right)\exp\left[-(t-s)\frac{3\alpha}{2m}-(t-s)\sqrt{\frac{\alpha^2}{4m^2}-2\omega^2}\right]\right)$$

The 2 in Eq.(61) cancels the ½ in Eq.(71) so that we may define $G(t-s)$ to be the right hand side of Eq.(71) sans the ½. This makes Eq.(61) look like

(72)
$$d_t \langle r^2 \rangle_t = \int_0^t ds\ G(t-s)\left(\frac{1}{m\beta}-\omega^2\langle r^2\rangle_s\right)$$

We are now in a position to use Laplace transforms, and especially the rules for Laplace transforms of convolution integrals and Laplace transforms of derivatives. Denote the Laplace transform variable conjugate to $t$ by $z$, and denote the Laplace transform of a fucntion of $t$, $f(t)$, by the same function of $z$ with a hat, $\hat{f}(z)$. The Laplace transform of Eq.(72) is

(73)
$$z\widehat{\langle r^2\rangle}_z - r_0^2 = \hat{G}(z)\left(\frac{1}{m\beta z}-\omega^2\widehat{\langle r^2\rangle}_z\right)$$

From the definition of $G(t-s)$ implied by Eq.(71) we find

(74)
$$\hat{G}(z) = \left(1+\frac{\alpha}{2m}\frac{1}{\sqrt{\frac{\alpha^2}{4m^2}-2\omega^2}}\right)\frac{1}{z+\frac{3\alpha}{2m}-\sqrt{\frac{\alpha^2}{4m^2}-2\omega^2}}+$$

$$\left(1-\frac{\alpha}{2m}\frac{1}{\sqrt{\frac{\alpha^2}{4m^2}-2\omega^2}}\right)\frac{1}{z+\frac{3\alpha}{2m}+\sqrt{\frac{\alpha^2}{4m^2}-2\omega^2}}$$

Simple algebra gives
(75)

$$\langle \widehat{r^2} \rangle_z = \frac{r_0^2 + \frac{\hat{G}(z)}{m\beta z}}{z + \omega^2 \hat{G}(z)}$$

Define $\mu$ by

(76)
$$\mu = \sqrt{\frac{\alpha^2}{4m^2} - 2\omega^2}$$

Therefore

(77)
$$\hat{G}(z) = \frac{\left(1 + \frac{\alpha}{2m\mu}\right)\left(z + \frac{3\alpha}{2m} + \mu\right) + \left(1 - \frac{\alpha}{2m\mu}\right)\left(z + \frac{3\alpha}{2m} - \mu\right)}{\left(z + \frac{3\alpha}{2m}\right)^2 - \frac{\alpha^2}{4m^2} + 2\omega^2}$$

$$= \frac{2\left(z + \frac{2\alpha}{m}\right)}{z^2 + 3\frac{\alpha}{m}z + \frac{2\alpha^2}{m^2} + 2\omega^2}$$

and

(78)
$$\frac{r_0^2 + \frac{\hat{G}(z)}{m\beta z}}{z + \omega^2 \hat{G}(z)} = \frac{r_0^2 \left(z^2 + 3\frac{\alpha}{m}z + \frac{2\alpha^2}{m^2} + 2\omega^2\right) + 2\left(z + \frac{2\alpha}{m}\right)\frac{1}{m\beta z}}{z^3 + \frac{3\alpha}{2m}z^2 + \left(\frac{2\alpha^2}{m^2} + 4\omega^2\right)z + 4\omega^2 \frac{\alpha}{m}}$$

$$= \frac{r_0^2(z - \lambda_+)(z - \lambda_-) + 2\left(z + 2\frac{\alpha}{m}\right)\frac{1}{m\beta z}}{\left(z + \frac{\alpha}{m}\right)(z - \gamma_+)(z - \gamma_-)}$$

where

$$\lambda_\pm \equiv -\frac{3\alpha}{2m} \pm \sqrt{\frac{\alpha^2}{m^2} - 8\omega^2}$$

and

$$\gamma_\pm \equiv -\frac{\alpha}{m} \pm \sqrt{\frac{\alpha^2}{m^2} - 4\omega^2}$$

Thus,

(79)

$$\widehat{\langle r^2\rangle}_z = \frac{r_0^2(z-\lambda_+)(z-\lambda_-) + \frac{2}{m\beta} + \frac{4\alpha}{m^2\beta z}}{\left(z+\frac{\alpha}{m}\right)(z-\gamma_+)(z-\gamma_-)}$$

We must now take the Laplace transform inverse of this quantity.

Computing the Laplace inverse of the right hand side of Eq.(79) makes use of several Laplace transform rules for $f(t)$ and its Laplace transform $\hat{f}(z)$:

(80)
$$\hat{f}(z) \leftrightarrow f(t)$$
$$z\hat{f}(z) - f(0) \leftrightarrow d_t f(t)$$
$$z^2\hat{f}(z) - zf(0) - d_t f(0) \leftrightarrow d_t^2 f(t)$$
$$\frac{1}{z}\hat{f}(z) \leftrightarrow \int_0^t ds\, f(s)$$

Simple algebra verifies

(81)
$$\frac{1}{\left(z+\frac{\alpha}{m}\right)(z-\gamma_+)(z-\gamma_-)} = \frac{1}{4\omega^2 - \frac{\alpha^2}{m^2}}\left(\frac{1}{z+\frac{\alpha}{m}} - \frac{1}{2}\left(\frac{1}{z-\gamma_+}\right) - \frac{1}{2}\left(\frac{1}{z-\gamma_-}\right)\right)$$

The inverse Laplace transform of this quantity is

(82)
$$S(t) \equiv \frac{1}{4\omega^2 - \frac{\alpha^2}{m^2}}\left(\exp\left[-\frac{\alpha}{m}t\right] - \frac{1}{2}\exp[\gamma_+ t] - \frac{1}{2}\exp[\gamma_- t]\right)$$

with the properties $S(0) = 0$ and $d_t S(0) = 0$. Using Eqs.(79-82), we get

(83)
$$\langle r^2\rangle_t = r_0^2\left(d_t^2 S + 3\frac{\alpha}{m}d_t S + \left(2\frac{\alpha^2}{m^2} + 2\omega^2\right)S\right) + \frac{2}{m\beta}S + \frac{4\alpha}{m^2\beta}\int_0^t ds\, S(s)$$

Although involving algebra only, the reduction of this expression takes quite a lot of manipulation. Begin with

(84)
$$\left(d_t^2 S + 3\frac{\alpha}{m}d_t S + \left(2\frac{\alpha^2}{m^2} + 2\omega^2\right)S\right) =$$

$$\frac{2\omega^2 \exp\left[-\frac{\alpha}{m}t\right] - \left(\frac{1}{2}\gamma_+^2 + \frac{3\alpha}{2m}\gamma_+ + \frac{\alpha^2}{m^2} + \omega^2\right)\exp[\gamma_+ t]}{4\omega^2 - \frac{\alpha^2}{m^2}}$$

$$-\frac{\left(\frac{1}{2}\gamma_-^2 + \frac{3\alpha}{2m}\gamma_- + \frac{\alpha^2}{m^2} + \omega^2\right)\exp[\gamma_- t]}{4\omega^2 - \frac{\alpha^2}{m^2}}$$

From Eq.(78) it follows that

(85)
$$\gamma_\pm^2 + 2\frac{\alpha}{m}\gamma_\pm + 4\omega^2 = 0$$

and

$$\frac{1}{2}\gamma_\pm^2 + \frac{3\alpha}{2m}\gamma_\pm + \frac{\alpha^2}{m^2} + \omega^2 = \frac{\alpha}{2m}\gamma_\pm + \frac{\alpha^2}{m^2} - \omega^2$$

Putting this into Eq.(84) and then that result into Eq.(83) yields

(86)
$$\langle r^2 \rangle_t = \frac{r_0^2}{4\omega^2 - \frac{\alpha^2}{m^2}}\left(2\omega^2 \exp\left[-\frac{\alpha}{m}t\right] - \left(\frac{\alpha}{2m}\gamma_+ + \frac{\alpha^2}{m^2} - \omega^2\right)\exp[\gamma_+ t]\right.$$

$$\left. - \left(\frac{\alpha}{2m}\gamma_- + \frac{\alpha^2}{m^2} - \omega^2\right)\exp[\gamma_- t]\right)$$

$$+ \frac{\frac{2}{m\beta}}{4\omega^2 - \frac{\alpha^2}{m^2}}\left(\exp\left[-\frac{\alpha}{m}t\right] - \frac{1}{2}\exp[\gamma_+ t] - \frac{1}{2}\exp[\gamma_- t]\right)$$

$$+ \frac{\frac{4\alpha}{m^2\beta}}{4\omega^2 - \frac{\alpha^2}{m^2}}\left(-\frac{m}{\alpha}\left(\exp\left[-\frac{\alpha}{m}t\right] - 1\right) - \frac{1}{2\gamma_+}(\exp[\gamma_+ t] - 1)\right.$$

$$\left. - \frac{1}{2\gamma_-}(\exp[\gamma_- t] - 1)\right)$$

Now note that

(87)
$$\frac{m}{\alpha} + \frac{1}{2\gamma_+} + \frac{1}{2\gamma_-} = \frac{2\frac{m}{\alpha}\gamma_+\gamma_- + \gamma_- + \gamma_+}{2\gamma_+\gamma_-} = \frac{\frac{m}{\alpha}\left(8\omega^2 - 2\frac{\alpha^2}{m^2}\right)}{8\omega^2}$$

$$= \frac{\frac{m}{\alpha}}{4\omega^2}\left(4\omega^2 - \frac{\alpha^2}{m^2}\right)$$

and define $\gamma$ by

(88)
$$\gamma \equiv \sqrt{\frac{\alpha^2}{m^2} - 4\omega^2}$$

Eq.(86) becomes

(89)
$$\langle r^2 \rangle_t = \frac{1}{m\omega^2 \beta}$$
$$-\frac{4\alpha}{m^2\beta}\left(\frac{1}{-\gamma^2}\right) exp\left[-\frac{\alpha}{m}t\right]\left(\frac{m}{\alpha} + \frac{1}{2\gamma_+}exp[\gamma t] + \frac{1}{2\gamma_-}exp[-\gamma t]\right)$$
$$+\frac{2}{m\beta}\left(\frac{1}{-\gamma^2}\right) exp\left[-\frac{\alpha}{m}t\right]\left(1 - \frac{1}{2}exp[\gamma t] - \frac{1}{2}exp[-\gamma t]\right)$$
$$+\frac{r_0^2}{(-\gamma^2)} exp\left[-\frac{\alpha}{m}t\right]\left(2\omega^2 - \left(\frac{\alpha}{2m}\gamma_+ + \frac{\alpha^2}{m^2} - \omega^2\right)exp[\gamma t]\right.$$
$$\left.- \left(\frac{\alpha}{2m}\gamma_- + \frac{\alpha^2}{m^2} - \omega^2\right)exp[-\gamma t]\right)$$
$$= \frac{1}{m\omega^2\beta}$$
$$+\frac{2}{m\beta}\left(\frac{1}{-\gamma^2}\right) exp\left[-\frac{\alpha}{m}t\right]\left(1 - 2 - \left(\frac{1}{2} + \frac{\alpha}{m\gamma_+}\right)exp[\gamma t] - \left(\frac{1}{2} + \frac{\alpha}{m\gamma_-}\right)exp[-\gamma t]\right)$$
$$+\frac{r_0^2}{(-\gamma^2)} exp\left[-\frac{\alpha}{m}t\right]\left(2\omega^2 - \left(\frac{\alpha}{2m}\gamma_+ + \frac{\alpha^2}{m^2} - \omega^2\right)exp[\gamma t]\right.$$
$$\left.- \left(\frac{\alpha}{2m}\gamma_- + \frac{\alpha^2}{m^2} - \omega^2\right)exp[-\gamma t]\right)$$
$$= \frac{1}{m\omega^2\beta}$$
$$+\left(r_0^2 - \frac{1}{m\omega^2\beta}\right) exp\left[-\frac{\alpha}{m}t\right]\left(\frac{2\omega^2}{-\gamma^2}\right) - exp\left[-\frac{\alpha}{m}t\right]\left(\frac{1}{-\gamma^2}\right) \times$$

$$\left\{ r_0^2 \left( \left( -\frac{\alpha^2}{2m^2} + \frac{\alpha\gamma}{2m} + \frac{\alpha^2}{m^2} - \omega^2 \right) exp[\gamma t] \right.\right.$$

$$+ \left( -\frac{\alpha^2}{2m^2} - \frac{\alpha\gamma}{2m} + \frac{\alpha^2}{m^2} - \omega^2 \right) exp[-\gamma t] \right)$$

$$\left.+ \frac{2}{m\beta} \left( \left( \frac{1}{2} + \frac{\alpha}{m\gamma_+} \right) exp[\gamma t] + \left( \frac{1}{2} + \frac{\alpha}{m\gamma_-} \right) exp[-\gamma t] \right) \right\}$$

The terms inside $\{...\}$ can be simplified using the identity

(90)
$$-\frac{\alpha^2}{2m^2} \pm \frac{\alpha\gamma}{2m} + \frac{\alpha^2}{m^2} - \omega^2 = \frac{\alpha}{2m}\gamma_\pm + \frac{\alpha^2}{m^2} - \omega^2$$

Therefore we get

(91)
$$\left\{ r_0^2 \left( \left( -\frac{\alpha^2}{2m^2} + \frac{\alpha\gamma}{2m} + \frac{\alpha^2}{m^2} - \omega^2 \right) exp[\gamma t] \right.\right.$$

$$+ \left( -\frac{\alpha^2}{2m^2} - \frac{\alpha\gamma}{2m} + \frac{\alpha^2}{m^2} - \omega^2 \right) exp[-\gamma t] \right)$$

$$\left.+ \frac{2}{m\beta} \left( \left( \frac{1}{2} + \frac{\alpha}{m\gamma_+} \right) exp[\gamma t] + \left( \frac{1}{2} + \frac{\alpha}{m\gamma_-} \right) exp[-\gamma t] \right) \right\}$$

$$= -\left( r_0^2 - \frac{1}{m\omega^2\beta} \right) 2\omega^2 \cosh(\gamma t) + r_0^2 \frac{\alpha^2}{m^2} \cosh(\gamma t) + r_0^2 \frac{\alpha\gamma}{m} \sinh(\gamma t)$$

$$+ \frac{2\alpha}{m^2\beta} \left( \frac{\gamma_- exp[\gamma t] + \gamma_+ exp[-\gamma t]}{4\omega^2} \right)$$

$$= \left( r_0^2 - \frac{1}{m\omega^2\beta} \right) \left( \frac{\alpha^2}{m^2} - 2\omega^2 \right) \cosh(\gamma t) - \left( r_0^2 - \frac{1}{m\omega^2\beta} \right) \frac{\alpha\gamma^2}{m\gamma} \sinh(\gamma t)$$

Finally, putting everything together, we get

(92)
$$\langle r^2 \rangle_t = \frac{1}{m\omega^2\beta} + \left( r_0^2 - \frac{1}{m\omega^2\beta} \right) exp\left[ -\frac{\alpha}{m}t \right] \times$$

$$\left(\left(\frac{2\omega^2}{-\gamma^2}\right) + \left(\frac{2\omega^2 - \frac{\alpha^2}{m^2}}{-\gamma^2}\right)\cosh(\gamma t) + \frac{\alpha}{m\gamma}\sinh(\gamma t)\right)$$

One more observation puts this formula into its final form

(93)
$$\left(\cosh\left(\frac{\gamma t}{2}\right) + \frac{\alpha}{m\gamma}\sinh\left(\frac{\gamma t}{2}\right)\right)^2$$
$$= \left(\left(\frac{2\omega^2}{-\gamma^2}\right) + \left(\frac{2\omega^2 - \frac{\alpha^2}{m^2}}{-\gamma^2}\right)\cosh(\gamma t) + \frac{\alpha}{m\gamma}\sinh(\gamma t)\right)$$

Therefore,

(94)
$$\langle r^2\rangle_t = \frac{1}{m\omega^2\beta} + \left(r_0^2 - \frac{1}{m\omega^2\beta}\right)exp\left[-\frac{\alpha}{m}t\right]\left(\cosh\left(\frac{\gamma t}{2}\right) + \frac{\alpha}{m\gamma}\sinh\left(\frac{\gamma t}{2}\right)\right)^2$$

and agrees with ([15] Chandrasekhar, Eq.(217)).

## VIII. Non-Markovian Distribution for the Harmonic Oscillator

Return to Eqs.(57-58) to see how the first moment of $r$ evolves in time. The initial conditions for Eq.(58) are $\langle r\rangle_0 = r_0$ and $d_t\langle r\rangle_0 = 0$ (see Eq.(57)). The solution is

(95)
$$\langle r\rangle_t = r_0 exp\left[-\frac{\alpha}{2m}t\right]\left(\cosh\left(\frac{\gamma t}{2}\right) + \frac{\alpha}{m\gamma}\sinh\left(\frac{\gamma t}{2}\right)\right)$$

Introduce the short-hand notation

(96)
$$T(t) \equiv exp\left[-\frac{\alpha}{2m}t\right]\left(\cosh\left(\frac{\gamma t}{2}\right) + \frac{\alpha}{m\gamma}\sinh\left(\frac{\gamma t}{2}\right)\right)$$

so that we can write

(97)
$$\langle r\rangle_t = r_0 T(t)$$
and

$$\langle r^2 \rangle_t = \frac{1}{m\omega^2 \beta} + \left(r_0^2 - \frac{1}{m\omega^2 \beta}\right) T(t)^2$$

Clearly $T(0) = 1$ and $T(\infty) = 0$ yielding the equipartition of energy result for $\langle r^2 \rangle_\infty$. The variance, $\sigma_t^2 \equiv \langle r^2 \rangle_t - (\langle r \rangle_t)^2$, is equal to

(98)
$$\sigma_t^2 = \frac{1}{m\omega^2 \beta}(1 - T(t)^2)$$

Because we have a Gaussian process we can immediately write down the conditional probability distribution

(99)
$$P_2(r, t; r_0, t_0) = \frac{1}{\sqrt{\frac{2\pi}{m\omega^2 \beta}(1 - T(t)^2)}} \exp\left[-\frac{(r - r_0 T(t))^2}{\frac{2}{m\omega^2 \beta}(1 - T(t)^2)}\right]$$

This is manifestly non-Markovian because of the $t$ dependence of $T(t)$. Note the two limits

(100)
$$\lim_{t \to t_0} P_2(r, t; r_0, t_0) = \delta(r - r_0)$$
$$\text{and}$$
$$\lim_{t \to \infty} P_2(r, t; r_0, t_0) = W_B(U(r))$$

where $W_B(U(r))$ is the normalized Boltzmann distribution for the potential energy $U(r) = \frac{1}{2} m\omega^2 r^2$. In the Gaussian Markovian case, the distribution $P_2$ could be used to construct all multi-time correlation functions ([21] Wang and Uhlenbeck, [11] Fox), but not in the non-Markovian case that we have here.

### IX. Non-equilibrium Thermodynamics

Instead of considering phase space trajectories and their time reversed partners, we have contracted the description and derived the time evolution in coordinate space. Time reversed phase space trajectories require the ability to reverse each and every particle momentum simultaneously and instantaneously. While this is possible in a molecular dynamics computation on a computer it is impossible in real experiments, and will always remain so. Alternatively we may ask if thermodynamic ideas can be extended into the non-equilibrium regime. For thermostated systems this means we should look at the Helmholtz free energy, $F$, that attains a minium value in equilibrium.

We will see that in the "over-damped" case ($\gamma\ real$) the Helmholtz free energy for a harmonic oscillator decreases monotonically in time. In the under-damped case ($\gamma\ imaginary$) monotonicity is broken and the Helmholtz free energy shows damped oscillations as it approaches its equilibrium value. The Helmholtz free energy is defined by

(101)
$$F \equiv E - TS$$

in which $E$ is the internal energy defined by

(102)
$$E \equiv \int dr\ U(r) P_2(r,t;r_0,t_0)$$

and $S$ is the entropy defined by

(103)
$$S \equiv -k_B \int dr\ ln\big(P_2(r,t;r_0,t_0)\big) P_2(r,t;r_0,t_0)$$

where $T$ is the Kelvin temperature, $k_B$ is Boltzmann's constant and the logarithm term could have a constant included to make its argument appropriately dimensionless but this would not show up in the time derivative and is, therefore, omitted. The time derivative of $F$ is

(104)
$$d_t F = \int dr\ \big(U(r) + k_B T\ ln\big(P_2(r,t;r_0,t_0)\big)\big)\, \partial_t P_2(r,t;r_0,t_0)$$

in which we have dropped the term

(105)
$$\int dr\ k_B T \frac{1}{P_2(r,t;r_0,t_0)} \big(\partial_t P_2(r,t;r_0,t_0)\big) P_2(r,t;r_0,t_0)$$
$$= \int dr\ k_B T \partial_t P_2(r,t;r_0,t_0) = 0$$

Eq.(105) follows from Eq.(44) because $P_2(r,t;r_0,t_0)$ here is $R(r,t)$ there and the right hand side of Eq.(44) begins with $b^\dagger$ which upon $r$ integration by parts vanishes since $R$ vanishes at the boundaries. In general, the non-Markovian nature

(convolution integral) of $\partial_t P_2(r,t;r_0,t_0)$ given by the right hand side of Eq.(44) makes this difficult to analyze. For insight, we return to our simple examples.

The first example is $U = 0$. The solution for $P_2(r,t;r_0,t_0)$ is Eq.(54). Eq.(104) in this case becomes

(106)
$$d_t F = \int dr \left(k_B T \ln(P_2(r,t;r_0,t_0))\right) \partial_t P_2(r,t;r_0,t_0)$$
$$= -k_B T \int dr \left(\frac{1}{2}\ln(4\pi D_E(t-t_0)) + \frac{(r-r_0)^2}{4D_E(t-t_0)}\right) \partial_t P_2(r,t;r_0,t_0)$$
$$= -k_B T \frac{d_t 2 D_E(t-t_0)}{4 D_E(t-t_0)}$$
$$= -k_B T \frac{1 - \exp\left[-(t-t_0)\frac{\alpha}{m}\right]}{2\left((t-t_0) + \frac{m}{\alpha}\left(\exp\left[-(t-t_0)\frac{\alpha}{m}\right] - 1\right)\right)}$$
$$\leq 0$$

This is a monotone decrease of the Helmholtz free energy and suggests that thermodynamics applies to all non-equilibrium states in this special case. The monotonicity follows from the positivity of the numerator and of the denominator of the fraction, for all times.

A more advanced example is the harmonic oscillator. In this case the internal energy is

(107)
$$E = \int dr \, \frac{1}{2} m\omega^2 r^2 \, P_2(r,t;r_0,t_0) = \frac{1}{2} m\omega^2 \langle r^2 \rangle_t$$
$$= \frac{1}{2\beta}(1 - T(t)^2) + \frac{1}{2} m\omega^2 r_0^2 T(t)^2$$

where we have used Eqs.(97-98). The entropy is given by
(108)

$$S = -k_B \int dr \, \ln(P_2(r,t;r_0,t_0))P_2(r,t;r_0,t_0)$$

$$= -k_B \int dr \left( -\frac{1}{2}\ln\left(\frac{2\pi}{m\omega^2\beta}(1-T(t)^2)\right) \right.$$

$$\left. -\frac{(r-r_0 T(t))^2}{\frac{2}{m\omega^2\beta}(1-T(t)^2)} \right) P_2(r,t;r_0,t_0)$$

$$= k_B \frac{1}{2}\ln\left(\frac{2\pi}{m\omega^2\beta}(1-T(t)^2)\right) + k_B \frac{\langle(r-r_0 T(t))^2\rangle}{\frac{2}{m\omega^2\beta}(1-T(t)^2)}$$

$$= \frac{k_B}{2}\left(1 + \ln\left(\frac{2\pi}{m\omega^2\beta}(1-T(t)^2)\right)\right)$$

Thus, for the harmonic oscillator,
(109)

$$F = \frac{1}{2\beta}(1-T(t)^2) + \frac{1}{2}m\omega^2 r_0^2 T(t)^2 - \frac{1}{2\beta}\left(1 + \ln\left(\frac{2\pi}{m\omega^2\beta}(1-T(t)^2)\right)\right)$$

Therefore, the time derivative is
(110)

$$d_t F = \left(\frac{1}{2\beta}\left(-1 + \frac{1}{(1-T(t)^2)}\right) + \frac{1}{2}m\omega^2 r_0^2\right) 2T(t)d_t T(t)$$

$$= \left(\frac{1}{\beta}\left(\frac{T(t)^2}{(1-T(t)^2)}\right) + m\omega^2 r_0^2\right) T(t)d_t T(t)$$

where $T(t)$ is defined in Eq.(96). The factor $\gamma$ was introduced in Eq.(88) and is real in the overdamped case, and is imaginary in the underdamped case. Consider the real case first. Introduce the parameter $\rho \equiv \left(\frac{2m\omega}{\alpha}\right)^2$. Overdamped case means $\rho < 1$ and underdamped case means $\rho > 1$. We can rewrite $T(t)$ in the form
(111)

$$T(t) = \frac{1}{2}\left(1 + \frac{1}{\sqrt{1-\rho}}\right)\exp\left[-\frac{\alpha}{2m}(1-\sqrt{1-\rho})t\right]$$
$$+ \frac{1}{2}\left(1 - \frac{1}{\sqrt{1-\rho}}\right)\exp\left[-\frac{\alpha}{2m}(1+\sqrt{1-\rho})t\right]$$

Elementary methods can be used to verify, in the overdamped case, $1 \geq T(t) \geq 0$. Therefore the sign of $d_t F$ in Eq.(110) is determined by the sign of $d_t T(t)$. Algebra can be used to easily verify

(112)
$$d_t T = -\frac{\alpha\rho}{4m\sqrt{1-\rho}}\exp\left[-\frac{\alpha}{2m}(1-\sqrt{1-\rho})t\right]\left(1 - \exp\left[-\frac{\alpha}{m}\sqrt{1-\rho}\,t\right]\right)$$
$$\leq 0$$

The Helmholtz free energy for the overdamped harmonic oscillator is monotone decreasing, and we can conclude that non-equilibrium thermodynamics applies far from equilibrium in this case.

The situation is fundamentally different in the underdamped harmonic oscillator case ($\rho > 1$ and $\gamma$ is imaginary). In this case $\gamma$ is expressed as

(113)
$$\gamma = i2\omega\sqrt{1 - \frac{1}{\rho}}$$

and the hyperbolic functions in Eq.(96) become trigonometric functions so that

(114)
$$T(t) = \exp\left[-\frac{\alpha}{2m}t\right]\left(\cos\left(\omega\sqrt{1-\frac{1}{\rho}}\,t\right) + \frac{1}{\sqrt{\rho-1}}\sin\left(\omega\sqrt{1-\frac{1}{\rho}}\,t\right)\right)$$
$$= \exp\left[-\frac{\alpha}{2m}t\right]\sqrt{\frac{\rho}{\rho-1}}\cos\left(\omega\sqrt{\frac{\rho-1}{\rho}}\,t - \theta\right)$$

where
$$\theta = \arctan\left(\frac{1}{\sqrt{\rho-1}}\right)$$

and

$$\cos(\theta) = \sqrt{\frac{\rho - 1}{\rho}}$$

Clearly, as $\rho$ goes from $1^+$ to $\infty$, $\theta$ goes from $\frac{\pi^-}{2}$ to 0 and $\cos(\theta)$ is positive for all of this. Thus at $t = 0$, $T(0)$ is positive, but then it shows damped oscillations during which $T(t)$ becomes negative repeatedly. For example as soon as $t$ increases so that $\left(\omega\sqrt{\frac{\rho-1}{\rho}} t - \theta\right)$ passes $\frac{\pi}{2}$, $T(t)$ becomes negative. In Eq.(110) the bracketed coefficient remains positive for all time because of the squares, but the factor of $T(t)d_t T(t)$ oscillates, partly because of the $T(t)$ oscillations just discussed, and partly from the $d_t T(t)$ oscillations to be discussed. From Eq.(114) we have

(115)
$$d_t T(t) = -\frac{\alpha}{2m} T(t) + \omega \exp\left[-\frac{\alpha}{2m} t\right] \sin\left(\omega \sqrt{\frac{\rho - 1}{\rho}} t - \theta\right)$$

This quantity also oscillates, although out of phase with $T(t)$. Using the first line of Eq.(114) one gets directly

(116)
$$d_t T(t) = -\frac{\alpha}{2m} T(t)$$
$$+ \exp\left[-\frac{\alpha}{2m} t\right]\left(-\omega\sqrt{\frac{\rho-1}{\rho}} \sin\left(\omega\sqrt{\frac{\rho-1}{\rho}} t\right) + \omega \frac{1}{\sqrt{\rho}} \cos\left(\omega\sqrt{\frac{\rho-1}{\rho}} t\right)\right)$$

$$= \exp\left[-\frac{\alpha}{2m} t\right]\left(\left(-\frac{\alpha}{2m} + \frac{\omega}{\sqrt{\rho}}\right)\cos\left(\omega\sqrt{\frac{\rho-1}{\rho}} t\right)\right.$$
$$+ \left(-\frac{\alpha}{2m}\frac{1}{\sqrt{\rho-1}} - \omega\sqrt{\frac{\rho-1}{\rho}}\right)\sin\left(\omega\sqrt{\frac{\rho-1}{\rho}} t\right)\right)$$

$$= -\omega \exp\left[-\frac{\alpha}{2m} t\right]\sqrt{\frac{\rho-1}{\rho}} \sin\left(\omega\sqrt{\frac{\rho-1}{\rho}} t\right)$$

The sign of the product $T(t)d_t T(t)$ is determined by the sign of

(117)
$$\sin\left(\omega\sqrt{\frac{\rho-1}{\rho}}t\right)\left(\cos\left(\omega\sqrt{\frac{\rho-1}{\rho}}t\right)+\frac{1}{\sqrt{\rho-1}}\sin\left(\omega\sqrt{\frac{\rho-1}{\rho}}t\right)\right)$$
$$=\frac{1}{2}\sin\left(2\omega\sqrt{\frac{\rho-1}{\rho}}t\right)+\frac{1}{\sqrt{\rho-1}}\sin^2\left(\omega\sqrt{\frac{\rho-1}{\rho}}t\right)$$

Note that the second term of the last line is positive for all $\rho > 1$. Only the first term can change the sign and this happens when the first term is more negative than the second term is positive. As it turns out, there is always a finite range of values of $t$ for which can make the expression in Eq.(117) negative for any $\rho > 1$. To see this note initially that if $\rho$ is very much greater than 1 then the first term dominates and periodically changes sign. The issue is whether this is possible for $\rho \to 1^+$. Look at the first line of Eq.(117) and assume $\rho = 1 + \frac{1}{\varphi^2}$ for large $\varphi$.

Changes of sign in Eq.(117) occur when (assuming $\sin\left(\omega\sqrt{\frac{\rho-1}{\rho}}t\right)$ does not vanish at the same time, a verifiably safe assumption).

(118)
$$\cos\left(\omega\sqrt{\frac{\rho-1}{\rho}}t\right)+\varphi\sin\left(\omega\sqrt{\frac{\rho-1}{\rho}}t\right)=0$$

This is equivalent with

(119)
$$\omega\sqrt{\frac{\rho-1}{\rho}}t = arctan\left(-\frac{1}{\varphi}\right)$$

which happens periodically in $t$. The amount of the $t$ axis for which the expression in Eq.(117) is negative decreases as $\varphi$ gets large. Negative values correspond with increases of the Helmholtz free energy. Therfore, in the case of the underdamped harmonic oscillator the generalization of thermodynamics to the non-equilibrium states fails because $F(t)$ is no longer monotonically decreasing.

## X. Underdamping in Sub-cellular Biology

Generally the molecular events occurring in sub-cellular biology are overdamped. An example is the elasticity of the neck linker of the kinesin molecule the attaches to the kinesin head and plays a central role in the kinesin mechanism ([8] Fox, chapter 4). The harmonic oscillator approach to the neck linker elasticity results in $\rho \sim 10^{-4}$ an extreme overdamped value. Not only is non-equilibrium thermodynamics reasonable for this regime, it is possible to approximate the dynamics by a Markovian dynamics for which the path integral techniques of Jarzinski and Crooks are valid (although here we are in coordinate space only, not full phase space).

The Markovian approximation to Eq.(43) can be justified with various degrees of rigor. Here, we will see how it comes about using the Dirac delta function because this approach is very transparent. Suppose you have the equation
(120)
$$\partial_t f(t) = \int_0^t ds \, exp[-\lambda(t-s)]g(s)$$
If $\lambda$ is large, write this equation in the form
(121)
$$\partial_t f(t) = \frac{2}{\lambda}\int_0^t ds \, \frac{\lambda}{2} exp[-\lambda|t-s|] \, g(s)$$
and contemplate the limit $\lambda \to \infty$. Because
(122)
$$\int_{-\infty}^{\infty} ds \, \frac{\lambda}{2} exp[-\lambda|t-s|] = 1$$
we can write
(123)
$$\lim_{\lambda \to \infty} \int_0^t ds \, \frac{\lambda}{2} exp[-\lambda|t-s|] \, g(s) = \int_0^t ds \, \delta(t-s) g(s) = \frac{1}{2} g(t)$$

where the factor of ½ arises from the end point rule for delta function integrals, i.e. $t$ is in the delta function and is the upper limit of integration. Thus, for large enough $\lambda$ it follows that
(124)

$$\partial_t f(t) = \frac{2}{\lambda} \int_0^t ds \, \frac{\lambda}{2} exp[-\lambda|t-s|] \, g(s) \to \frac{1}{\lambda} g(t)$$

If we apply this result to Eq.(43), we see that the limits of integration inside the time ordered exponential become the same, i.e. $t = s$ so that the time ordered exponential simply becomes 1. Thus the Markovian approximation to Eq.(43) is

(125)
$$\partial_t R(r,t) = -\frac{m}{\alpha} b^\dagger b R(r,t) = \frac{1}{\alpha\beta} \partial_r (\partial_r + \beta U') R(r,t)$$

wherein we have used Eq.(27) for general potential energy $U$, and the constant in front of the right hand side is Einstein's formula for the diffusion constant. Eq.(125) may be recognized as diffusion in a potential and is sometimes called the Smoluchowski equation ([15] Chandrasekhar, pp. 40-41). Because the Markov approximation requires that $\lambda$ is large (the relaxation time is very short) this approximation only holds for the overdamped case.

Using the last line of Eq.(9) we can show that the Helmholtz free energy given by analogs of Eqs.(101-103) satisfies

(126)
$$d_t F = \int dr \, \big(U(r) + k_B T \, ln(R(r,t))\big) \partial_t R(r,t)$$
$$= \int dr \, \big(U(r) + k_B T \, ln(R(r,t))\big) D_E \partial_r (\partial_r + \beta U') R(r,t)$$
$$= -\int dr \, \Big[\partial_r \big(U(r) + k_B T \, ln(R(r,t))\big)\Big] D_E (\partial_r + \beta U') R(r,t)$$
$$= -D_E \int dr \, \Big(U' + k_B T \frac{1}{R(r,t)} R'(r,t)\Big) \big(R'(r,t) + \beta U' R(r,t)\big)$$
$$= -D_E \int dr \, \frac{k_B T}{R(r,t)} \big(R'(r,t) + \beta U' R(r,t)\big)^2 \leq 0$$

The third line follows from integration by parts and the vanishing of $R$ on the boundary. Thus a non-equilibrium thermodynamics is valid for all non-equilibrium states in the Markov approximation and the Helmholtz free energy decreases monotonically in an undriven thermostated system. This situation in no longer valid in the full non-Markovian picture (Eq.(43)) as we saw for the underdamped harmonic oscillator and as would be true for underdamped motion in a general potential.

Underdamped motion in sub-cellular biology is rarely documented because this realm is seriously overdamped in most situations. In the example of kinesin referred to earlier the elasticity time scale was found to be nanoseconds, whereas the damping time scale was picoseconds. This means the kinesin dynamics is very overdamped. Recently there have been claims that some protein-ligand interactions are indeed underdamped ([24] Turton et al.). In these studies the elasticity time scale is picoseconds and because the proteins fragments of the type studied are much larger than the kinesin heads the relaxation time scale is much less than picoseconds. These circumstances make the system underdamped and, therefore, non-Markovian. I quote from the abstract of Turton et al.:

"Low-frequency collective vibrational modes in proteins have been proposed as being responsible for efficiently directing biochemical reactions and biological energy transport. However, evidence of the existence of delocalized vibrational modes is scarce and proof of their involvement in biological function absent. Here we apply extremely sensitive femtosecond optical Kerr-effect spectroscopy to study the depolarized Raman spectra of lysozyme and its complex with the inhibitor triacetylchitotriose in solution. Underdamped delocalized vibrational modes in the terahertz frequency domain are identified and shown to blue-shift and strengthen upon inhibitor binding. This demonstrates that the ligand-binding coordinate in proteins is underdamped and not simply solvent-controlled as previously assumed. The presence of such underdamped delocalized modes in proteins may have significant implications for the understanding of the efficiency of ligand binding and protein–molecule interactions, and has wider implications for biochemical reactivity and biological function."

Thus the significance of underdamped motion in sub-cellular biology may grow in the future. Its description is intrinsically non-Markovian and not controlled by thermodynamic rules. In particular the Helmholtz free energy does not decrease monotonically. Perhaps the content of this paper will help in the application of physically realistic methods to the understanding of thermostated systems in nano-biology and in nano-technology.

17 September 2014

Ronald F. Fox
Smyrna, Georgia